\documentclass[12pt]{article}
\usepackage[intlimits,tbtags]{amsmath}
\usepackage{amsfonts,amssymb,amsthm}
\usepackage[latin1]{inputenc}
\usepackage[usenames,dvips]{color}

\def\beq{\begin{equation}}
\def\eeq{\end{equation}}
\def\bea{\begin{eqnarray}}
\def\eea{\end{eqnarray}}
\def\nn{\nonumber}

\begin{document}

\topmargin -0.5cm
\oddsidemargin -0.8cm
\evensidemargin -0.8cm
\pagestyle{empty}

\begin{center}
\vspace*{5mm}
{\bf From Quantum Affine Symmetry to Boundary Askey-Wilson Algebra and Reflection Equation} \\
\vspace*{0.4cm}
{\bf B.Aneva${}^{a}$ , M.Chaichian${}^{b}$ , P.P.Kulish${}^{c}$ } \\
\vspace*{0.2cm}
${}^{a}$ {INRNE, Bulgarian Academy of Sciences, 1784 Sofia, Bulgaria}\\
\vspace*{0.3cm}
${}^{b}$ {High Energy Physics Division, Department of Physical Sciences, University of
Helsinki and  Helsinki Institute of Physics, P.O.Box 64, 00014 Helsinki, Finland}\\
\vspace*{0.3cm}
${}^{c}$ {St.Petersburg Department of Steklov Institute of Mathematics, Fontanka, 27,
St.Petersburg, 191023, Russia
}\\
\vspace*{0.6cm}
{\bf Abstract} \\
\end{center}
Within the quantum affine algebra representation theory we
construct linear covariant operators that generate the Askey-Wilson
algebra. It has the property of a coideal subalgebra, which can be interpreted
as the boundary symmetry algebra of  a model with
quantum affine symmetry in the bulk. The generators of the  Askey-Wilson
algebra are implemented to construct an operator valued $K$- matrix, a solution of a
spectral dependent reflection equation. We consider the open driven
diffusive system where the Askey-Wilson algebra arises as a boundary
symmetry and can be  used for an exact solution of the model in the stationary state.
We discuss the possibility of a solution beyond the stationary state on the
basis of the proposed relation of the Askey-Wilson algebra to the reflection equation.

 PACS numbers 02.30.Ik, 11.30.Na, 05.50.+k, 05.70.Ln

\setcounter{page}{1}
\pagestyle{plain}

\section{Introduction}
Quantum affine symmetries \cite {kul1} - \cite {dem} are intensively developed as
rich mathematical
structures linking various branches of mathematical physics, like topological
quantum field theories, integrable lattice models of statistical physics,
rational conformal theories. They are implemented with the ultimate goal to
understand and explore the consequences of the symmetries for description of
physical systems.

Our work concerns the application of quantum symmetries for the exact solvability
of many particle lattice systems interacting with stochastic dynamics.

The main idea of integrability of lattice systems (within the
inverse scattering method \cite {skl1}) is the existence of a
family of commuting transfer matrices, depending on a spectral
parameter. For quantum spin chains the transfer matrices give rise
to infinitely many mutually commuting conservation laws. This is
the abelian symmetry of the system. The infinitely many commuting
conserved charges can be diagonalized simultaneously and their
common eigenspace is finite dimensional in most cases. Thus the
abelian symmetry reduces the degeneracies of the spectrum from
infinite to finite which is the reason for integrability. In
addition many systems possess nonabelian symmetries. They
determine the $R$ matrix operator, a solution of the Yang-Baxter
equation, up to an overall scalar factor and are identified as the
quantum bulk symmetries. In the presence of general boundaries the
quantum symmetry, and the integrability of the model as well, are
broken. However with suitably chosen boundary conditions \cite
{skl2, kul2} a remnant of the bulk symmetry may survive and the
system possesses hidden boundary symmetries, which determine a
$K$-matrix, a solution of a boundary Yang-Baxter equation and
allow for the exact solvability. Such non local boundary symmetry
charges were originally obtained for the sine Gordon model
\cite{nep} and generalized to affine Toda field theories
\cite{del}, and derived from spin chain point of view as commuting
with the transfer matrix for a special choice of the boundary
conditions \cite{doi} or analogously as the one boundary
Temperley-Lieb algebra centralizer in the "nondiagonal" spin $1/2$
representation \cite{nic}.

In this paper we consider an algebraic prescription to construct
two operators, possessing a coideal property with respect to the
quantum affine $U_q(\hat {sl}(2))$. We show that these operators
generate an Askey-Wilson algebra which thus turns to be a coideal
subalgebra of the quantum affine $U_q(\hat {sl}(2))$. (See also
\cite {ter1} and \cite {bas1, bas3} for previous discussions of
these problems.) We argue that one can construct a $K$-matrix in
terms of the Askey-Wilson algebra generators, which satisfies a
boundary Yang-Baxter equation (known as a reflection  equation).
As an example of
an Askey-Wilson boundary symmetry we consider a model of
nonequilibrium physics, the open asymmetric exclusion process with
most general boundary conditions. This model is exactly solvable
in the stationary state within the matrix product ansatz to
stochastic dynamics and it can be shown that  the boundary
operators generate the Askey-Wilson algebra. The model is
equivalent to the integrable spin $1/2$ XXZ chain with most
general boundary terms, whose bulk Hamiltonian (infinite chain)
possesses the quantum affine symmetry $U_q(\hat {sl}(2))$.

\section{The Quantum Affine $U_q(\hat {sl}(2))$}
In this section we recall the definition of the affine $U_q(\hat {sl}(2))$
\cite {jim, drin, pres}.
We fix a real number $0<q<1$ (in the general case $q$ is complex)
and we use the $q$-symbol in the form
\beq
  [x]=\frac {q^{x/2}-q^{-x/2}}{q^{1/2}-q^{-1/2}}\equiv [x]_{q^{1/2}}
\label{1}
\eeq
The quantum affine $U_q(\hat {sl}(2))$ is defined as the
associative algebra with a unit
with generators $E^{\pm}_i$ and $q^{H_i}$,
$i=0,1$ in the Chevalley basis and defining relations
\bea
    q^{H_i}q^{-H_i}&=&q^{-H_i}q^{H_i}=1 \\  \nn
      q^{H_0}q^{H_1}&=&q^{H_0+H_1}=q^c
\label{2}
\eea
\bea
      q^{H_i}E_i^{\pm}q^{-H_i}&=&q^{\pm 1}E_i^{\pm}    \\  \nn
      q^{H_i}E_j^{\pm}q^{-H_i}&=&q^{\mp 1}E_j^{\pm}   \\  \nn
      [E^+_i, E^-_j]&=&
    \delta_{ij} \frac {q^{H_i}-q^{-H_i}}{q^{1/2}-q^{-1/2}}
\label{3}
\eea
together with the $q$-Serre relations
\beq
  (E_i^{\pm})^3E_j^{\pm}-[3](E_i^{\pm})^2E_j^{\pm}E_i^{\pm}
       +[3]E_i^{\pm}E_j^{\pm}(E_i^{\pm})^2 - E_j^{\pm}(E_i^{\pm})^3 = 0, \quad i\neq j
\label{3} \eeq where $[3]=1+q+q^{-1}$. The element $c=H_0+H_1$ is
central and its value is the level of the affine $U_q(\hat
{sl_2})$. The algebra is endowed with the structure of a Hopf
algebra. Namely, the coproduct $\Delta$, the counit $\epsilon$ and
the antipode $S$ are defined as \bea \Delta (E^+_i)&=&E^+_i\otimes
q^{-H_i/2}+q^{H_i/2}\otimes E^+_i   \\  \nn \Delta
(E^-_i)&=&E^-_i\otimes q^{-H_i/2}+q^{H_i/2}\otimes E^-_i   \\  \nn
\Delta (H_i)&=&H_i\otimes I + I\otimes H_i \label{4} \eea \beq
 \epsilon (E^+_i) = \epsilon (E^-_i) = \epsilon (H_i)=0, \quad\quad \epsilon (I) =1
\label{5}
\eeq
 \beq
S(E^{\pm}_i)=-q^{\mp 1/2}E^{\pm}_i,  \quad  S(H_i)=-H_i, \quad
S(I)= 1 \label{6} \eeq Let $\alpha_0$ denote the longest root and
$\rho$ is $1/2$ the sum of positive roots. We consider the
$U_q(\hat {sl}(2))$ algebra with a scaling element $d$, defined by
$(d, \alpha_0)=1$ ( $(,)$ is  the nondependence bilinear form on
the Cartan subalgebra) and denote
$h=(\alpha_0,\alpha_0)+2(\rho,\alpha_0)$. With a
finite-dimensional representation $\pi_V$ of $U_q(\hat {sl}(2))$
one associates the quantum $R$-matrix $R^{VV}(\lambda)$ which acts
 in $V\otimes V$ and
satisfies the Yang-Baxter equation.
One has also $[d\otimes d,R]=0$. The universal $R$-matrix $R(\lambda)$
is uniquely defined \cite {resh} by the first terms in its expansion in powers
of the Chevalley generators of $U_q(\hat {sl}(2))$
\beq
R=q^{c\otimes d+d\otimes c + \sum_{i=0}^1H_i\otimes H^i}
\left (1\otimes 1 +
(q^{1/2}-q^{-1/2})\sum_{i=0}^1q^{H_i/2}E_i^+\otimes q^{-H_i/2}E_i^- +...\right )
\label{}
\eeq
One can define an automorphism
$T_{\lambda}$
\beq
 T_{\lambda}H_i=H_i, \quad\quad
T_{\lambda}E_i^{\pm}=\lambda^{\pm 1}E_i^{\pm}, \quad\quad i=0,1
\label{}
\eeq
and put $R(\lambda)=(T(\lambda)\otimes id)R$. For a fixed
finite-dimensional representation $(\pi, V)$ of the quotient algebra
$U_q(\hat {sl}(2))$, obtained by setting $c=0$,  one has
$R^{VV}(\lambda)=(\pi\otimes \pi)R(\lambda)$. Following \cite {resh},
one introduces the  currents
$L^{\pm}(\lambda )\in End V \otimes U_q(\hat {sl}(2))$, given by
$L^+=(id\otimes \pi_V)(R), L^-=(id\otimes \pi_V)(R^t)$, which are
 explicitly expressed
in terms of the Chevalley generators of $U_q(\hat {sl}(2))$. The
operators $L^{\pm}(\lambda)$ generate a Hopf algebra $A(R)$ and
their matrix coefficients generate an algebra $A_0(R)\subset
A(R)$. Let $L(\lambda)$ denote the quantum current \beq
L(\lambda)= L^+(\lambda q)(L^-(\lambda)^{-1} \label{} \eeq with a
finite Laurent series expansion \beq L(\lambda)= \sum _n
l_V(n)\lambda ^{-n-2}, \quad\quad n\in Z \label{} \eeq A theorem
(by Reshetikhin and Semenov-Tian-Shansky \cite {resh}) states that
the element $t(\lambda)=tr_q(L(\lambda))$ lies in the center of
the quotient algebra of $A(R)$, obtained by setting $c=-h$. Hence
from the explicit expressions of the currents $L^{\pm}$ in terms
of the Chevalley generators follows that $t(\lambda)$  is the
generating function of the Casimir elements of the quotient
algebra of $U_q(\hat {sl}(2))$, obtained by setting $c=-h$.

For our purposes we will need a
slightly different realization of the algebra in terms of the Chevalley generators
and following \cite{ter1} we define a new
basis in $U_q(\hat {sl}(2))$ generated by $H_i, Q^s_i, \bar Q_i^s$
\beq
  Q^s_i=(q^{1/2}-q^{-1/2})E_i^+q^{-H_i/2}+q^{-H_i}, \quad\quad\quad
\bar Q^s_i =-(q^{1/2}-q^{-1/2})E_i^-q^{-H_i/2} + q^{-H_i}
\label{8}
\eeq
Let now $u, u^*, v, v^*$ be some (complex) scalars.
We denote
\beq
U=uQ^s_0, \qquad U^*=u^*Q^s_1, \qquad V=v\bar Q^s_0, \qquad V^*=v^*\bar Q^s_1
\label{13}
\eeq
Then we have
\beq
  q^{1/2}VU-q^{-1/2}UV=(q^{1/2}-q^{-1/2})vu
\label{14}
\eeq
\beq
  q^{1/2}UV^*-q^{-1/2}V^*U=(q^{1/2}-q^{-1/2})uv^*q^{-c}
\label{15}
\eeq
\beq
  q^{1/2}V^*U^*-q^{-1/2}U^*V^*=(q^{1/2}-q^{-1/2})v^*u^*
\label{16}
\eeq
\beq
  q^{1/2}U^*V-q^{-1/2}VU^*=(q^{1/2}-q^{-1/2})u^*vq^{-c}
\label{17}
\eeq
The operators $U, U^*, V,V^*$ satisfy the following $q$-Serre relations which are
the direct consequence of (4)
\beq
U^3U^* - [3]U^2U^*U + [3]UU^*U^2 +U^*U^3 =0
\label{18}
\eeq
\beq
U^{*3}U - [3]U^{*2}UU^* + [3]U^*UU^{*2} +UU^{*3} =0
\label{19}
\eeq
\beq
V^3V^* - [3]V^2V^*V + [3]VV^*V^2 +V^*V^3 =0
\label{20}
\eeq
\beq
V^{*3}V - [3]V^{*2}VV^* + [3]V^*VV^{*2} +VV^{*3} =0
\label{21}
\eeq
We can now consider the linear combinations
\beq
    A=U+V, \quad\quad\quad  A^*=U^*+V^*
\label{22} \eeq It can be verified directly by using the
$q$-commutation relations (14-17) between the operators
$U,V,U^*,V^*$ and the $q$-Serre relations (18-21) that the
operators $A,A^*$ satisfy the following relations which are the
defining relations of a tridiagonal   algebra \bea A^3A^*
- [3]_qA^2A^*A + [3]_qAA^*A^2 +A^*A^3 &=&-uv(q-q^{-1})^2[A,A^*] \\
\nn A^{*3}A - [3]_qA^{*2}AA^* + [3]_qA^*AA^{*2}
+AA^{*3}&=&-u^*v^*(q-q^{-1})^2[A^*,A] \label{24} \eea where
$[3]_q=q+q^{-1}+1$ and $[X,Y]=XY-YX$. In the following section we
are going to consider a  general realization of the Askey-Wilson
(AW) algebra as a coideal subalgebra of the quantum affine
$U_q(\hat {sl}(2))$ and the related tridiagonal algebra.

\section{The  Askey-Wilson and tridiagonal algebras}

We begin this section with some definitions \cite {ter2, ter3,
ter4, itate}. Let $V$ be a vector space with (in)finite positive
dimension. A tridiagonal pair on $V$ is an ordered pair $A, A^*$
where $A: V\rightarrow V$ and $A^*: V\rightarrow V$ are linear
transformations (see Definition 1.1 on p.2 in \cite {itate}) and
is a Leonard pair if the conditions hold: 1. There exists a basis
for $V$ with respect to which the matrix representing $A$ is
diagonal, and the matrix representing $A^*$ is irreducible
tridiagonal. 2. There exists a basis for $V$ with respect to which
the matrix representing $A^*$ is diagonal and the matrix
representing $A$  is irreducible tridiagonal. (A tridiagonal
matrix is  irreducible whenever all entries on the superdiagonal
and subdiagonal are nonzero.) A tridiagonal pair $A,A^*$ on $V$ is
a  Leonard pair
 if for each $A,A^*$  all eigenspaces are of dimension $1$.\\
{\it Definition 1}: For a Leonard pair $A,A^*$  on $V$
there exists a sequence of scalars
 $\beta, \gamma, \gamma^*, \rho, \rho^*, \omega, \eta, \eta^*$,
such that
\bea
-\beta AA^*A + A^2A^* +A^*A^2 - \gamma \{A,A^*\}-\rho A^* &=&
 \gamma^*A^2 +\omega A + \eta  \\ \nn
-\beta A^*AA^* + A^{*2}A+AA^{*2}- \gamma^* \{A,A^*\}-\rho^*A&=&
 \gamma A^{*2} +\omega A^* +\eta^*
\label{32}
\eea
which is uniquely determined by the pair.
The equations (24) are called the Askey-Wilson relations.\\
{\it Definition 2}: For a tridiagonal pair $A,A^*$ on $V$
there exists a sequence of
scalars $\beta, \gamma, \gamma^*, \rho, \rho^*$,
such that
\bea
 [A, -\beta AA^*A + A^2A^* +A^*A^2 -\gamma \{A,A^*\}-\rho A^*] &=&0  \\ \nn
 [A^*,-\beta A^*AA^* + A^{*2}A+AA^{*2}-\gamma^* \{A,A^*\}-\rho^*A] &=& 0
\label{32} \eea which is   uniquely determined by the pair, if the
dimension of $V$ is at least four. It has been proved \cite{vid}
 for a Leonard pair that the two definitions are equivalent.\\
{\it Definition 3}: A tridiagonal algebra, respectively the
Askey-Wilson algebra, is an associative algebra with a unit
generated by the pair $A,A^*$ subject to the relations (25),
respectively (24).

 Affine transformations
\beq
 A \rightarrow tA+c', \qquad A^* \rightarrow t^*A^*+c^*
\label{31}
\eeq
where $t, t^*, c',c^*$ are some scalars, act on the operators
$A,A^*$ of the Askey-Wilson relations and  can be used to bring a
tridiagonal pair in a reduced form with $\gamma=\gamma^*=0$.
(The label $c'$ is used to avoid confusion with
 the $U_q(\hat {sl}(2))$ central element.)
Examples are the $q$-Serre relations  with
 $\beta =q+q^{-1}$ and $\gamma = \gamma^* = \rho = \rho^* =0$
and the Dolan-Grady relations \cite {dol} with
$\beta=2, \gamma = \gamma^* =0, \rho = k^2, \rho^* =k^{*2}$
\bea
[A,[A,[A,A^*]]]=k^2[A,A^*] \\  \nn
[A^*,[A^*,[A^*,A]]]=k^{*2}[A^*,A]
\label{46}
\eea

The algebra (24) was first considered by Zhedanov \cite {zhe1} who
showed that  the Askey-Wilson polynomials give raise to two
infinite-dimensional matrices satisfying the AW relations. The
tridiagonal relations have recently been discussed in a more
general framework \cite {ter2, ter3, koor} and a particular case
of a tridiagonal algebra with $\rho = \rho^*$ through a
homomorphism to $U_q(\hat{sl}(2))$ was proposed in \cite {bas3}.
Leonard pairs have been classified according to their dependence
on the sequence of scalars \cite {ter2, ter3} and a correspondence
to the orthogonal polynomials in the Askey-Wilson scheme was
given. In \cite {koor} the AW algebra (24) with
$\gamma=\gamma^*=0$ has been equivalently described as an algebra
with two generators and with structure constants determined in
terms of the elementary symmetric polynomials in four parameters
$a,b,c,d, abcd \neq q^m, m=0,1,2...; q\neq 0, q^k\neq 1,
k=1,2,...$.

We can now formulate the following statement which defines a
homomorphism of the AW algebra to the quantized affine algebra
$U_q(\hat {sl}(2))$ (see \cite {bas3} for the relation to
tridiagonal algebra).

{\it Proposition I}: Let $u,u^*,v,v^*,k,k^*$ be some scalars.
The operators $A, A^*$ defined by
\bea
A&=&uE_0^+q^{-H_0/2}+vE_0^-q^{-H_0/2}+kq^{-H_0}  \\  \nn
 A^*&=&u^*E_1^+q^{-H_1/2}+v^*E_1^-q^{-H_1/2}+k^*q^{-H_1}
\label{24}
\eea
(It is assumed that $E_i^{\pm}$ in (28) are rescaled by $\pm (q^{1/2}- q^{-1/2})$
according to (12))
and their $q$-commutator
\beq
 [A,A^*]_q=q^{1/2}AA^*-q^{-1/2}A^*A
\label{25}
\eeq
form a closed linear algebra, the Askey-Wilson algebra
\bea
[[A,A^*]_q,A]_q = -\rho A^* -\omega A - \eta  \\ \nn
[A^*,[A,A^*]_q]_q = -\rho^*A -\omega A^* -\eta^*
\label{26}
\eea
where
the (representation dependent) structure constants are given by
\beq
 -\rho= uv(q-q^{-1})^2, \quad
-\rho^*= u^*v^*(q -q^{-1})^2
\label{27}
\eeq
\beq
-\omega= -(q^{1/2}-q^{-1/2})^2\left (kk^*+l_V^0(uu^*q^{1/2}+v^*vq^{-1/2})\right )
\label{28}
\eeq
\beq
-\eta= (q-q^{-1})(q^{1/2}-q^{-1/2})\left ( -k(uu^*q^{1/2}+v^*vq^{-1/2})
  -l_V^0uvk^*\right )
\label{29}
\eeq
\beq
-\eta^*= (q-q^{-1})(q^{1/2}-q^{-1/2})\left ( -k^*(uu^*q^{1/2}+v^*vq^{-1/2})
  -l_V^0u^*v^*k\right )
\label{30} \eeq The element $l_V^0$ is  the coefficient $l_V(n)$
to $\lambda ^{-n-2}$, for $n=0$, in the Laurent series (11) of the
quantum current for any highest weight module over  $U_q(\hat
{sl}(2))$ and is central (the quadratic Casimir element) in the
quotient algebra of $U_q(\hat {sl}(2))$, obtained by setting
$c=-h$ (with $h=1$). On a highest weight irreducible
representation $V$ $l_V^0$ is a scalar.

We note that eq.(28) defines the
homomorphism of the AW algebra to the affine algebra $U_q(\hat
{sl}(2))$. For $k=u+v, k^*=u^*+v^*$ one recovers the particular
case (22). Special cases of this homomorphism is the
representation considered by Terwilliger \cite{ter2} with
$\rho=\rho^*=0$ and the one by Baseilhac \cite{bas1, bas3} and
Baseilhac and Koizumi \cite{bas} with $u=v^*$ and $v=u^*$ in (28).
Making use of the evaluation representation for the $U_q(\hat
{sl}(2))$ generators in (28) \beq \pi_{\nu} (E_1^{\pm})= E^{\pm},
\quad \pi_{\nu} (E_0^{\pm})=\nu^{\pm 1}E^{\mp}, \quad \pi_{\nu}
(q^{H_1})=q^H, \quad   \pi_{\nu} (q^{H_0})=q^{-H} \label{47} \eeq
where $E^{\pm}, H$ are the $U_q(sl(2))$ generators, we obtain the
Granovskii and Zhedanov realization \cite{zhe2}.

The algebraic relations (23) follow from (30) by taking the
commutator with $A$ and $A^*$ respectively.  The homomorphism (28)
defines the AW algebra with two generators as the linear
covariance algebra for the $U_q(\hat {sl}(2))$ with operator
valued structure constants. The tridiagonal algebra (23), denoted
TD, is obtained through the chain of homomorphisms $TD \rightarrow
AW \rightarrow U_q(\hat {sl}(2))$.

{\it Definition 4}: The AW algebra with two generators $A, A^*$
defined by the homomorphism  (28) as a coideal subalgebra yields a
deformation in two parameters $\rho, \rho^*$ of the $q$-Serre
relations of level zero  quantum affine $U_q(\hat {sl}(2))$, such
that it results  in  a shift of the central charge to a non zero
value $c=-1$.

Taking the limit $q\rightarrow 1$ in the defining relations (23) one obtains
a two parameter deformation of the Serre relations of level zero affine
$sl(2)$, known as the Dolan-Grady relations.

From the explicit realization of the operators $A,A^*$ it follows that they
generate a linear covariance algebra for the
 $U_q(\hat {sl}(2))$ which has the property of a coideal subalgebra. Let
$B_q(\hat {sl}(2))$  denote the algebra generated by $A, A^*$.

{\it Proposition II}: The Askey-Wilson algebra defined by the homomorphism (28)
is a coideal subalgebra of  $U_q(\hat {sl}(2))$. The proof is straightforward
by using the comultiplication (5). One has
\bea
\Delta (A)=   I \otimes A +   (A - kI)  \otimes   q^{-H_0}          \\  \nn
\Delta (A^*)=  I \otimes A^* + (A^*-k^*I)\otimes q^{-H_1}
\label{48}
\eea
where the expressions on the RHS of  (36) obviously belong to
  $B_q(\hat {sl}(2))\otimes U_q(\hat {sl}(2))$.

\section{Representations of the Askey-Wilson algebra}

The Askey-Wilson algebra is known to possess very important properties which
allow to obtain its ladder representations.
We briefly sketch these properties (for details
see \cite {ter2, zhe1, koor}).
Namely, there is a representation with basis $f_r$
with respect to which the operator $A$ is diagonal
\beq
 Af_r=\lambda _r f_r
\label{48}
\eeq
where the eigenvalues satisfy a quadratic equation
\beq
 \lambda_r^2 +\lambda_s^2 -(q+q^{-1})\lambda_r \lambda_s -\rho=0,
\label{49}
\eeq
(which yields two different eigenvalues $\lambda_{r+1}$ and $\lambda_{r-1}$
for a fixed eigenvalue $\lambda_r$)
and the operator $A^*$ is tridiagonal
\beq
 A^*f_r=a_{r+1}f_{r+1}+b_rf_r+c_{r-1}f_{r-1}
\label{50}
\eeq
Depending on the sign of $\rho$ the spectrum of the diagonal operator
is hyperbolic of the
form $sh$ or $ch$ and $exp$ if $\rho =0$.
The algebra possesses a duality property. Due to the duality property
the dual basis exists with respect to  which
the operator $A^*$ is diagonal and the operator $A$ is tridiagonal.
We have
\bea
 A^*f_p^*=\lambda^*_pf^*_p \\  \nn
 Af^*_s=a^*_{s+1}f^*_{s+1}+b^*_sf^*_s +c^*_{s-1}f^*_{s-1}
\label{51} \eea where $\lambda^*_p$ satisfies the quadratic
equation (38) with $-\rho$ replaced by $-\rho^*$. The overlap
function of the  two basis $\langle s \vert r \rangle = \langle
f^*_s \vert f_r \rangle$ can be expressed in terms of the
Askey-Wilson polynomials. Let $p_n=p_n(x;a,b,c,d)$ denote the
$n$-th Askey-Wilson polynomial \cite {ask} depending on four
parameters $a,b,c,d$ \beq p_n(x;a,b,c,d)= _4\Phi_3 \left ( \begin
{array}{c}
   q^{-n},abcdq^{n-1},ay,ay^{-1} \\
     ab,ac,ad \end{array} \vert q;q \right )
 \label{52}
\eeq with $p_0=1$, $x=y+y^{-1}$ and $0<q<1$. Then there is a basic
representation of the AW algebra \cite {koor}  which is a
representation of the tridiagonal algebra as well \cite {ter3}
$$
[A, A^2A^* -(q +q^{-1})AA^*A+A^*A^2 +abcdq^{-1}(q-q^{-1})^2A^*]=0  $$
$$
[A^*,A^{*2}A -(q+q^{-1})A^*AA^* +AA^{*2}+(q-q^{-1})^2A]=0
$$
in the space of symmetric Laurent polynomials $f[y]=f[y^{-1}]$ with a
basis $(p_0,p_1,...)$
as follows
\beq
   Af[y]=(y+y^{-1})f[y], \qquad    A^*f[y]= \mathcal{D}f[y]
\label{53}
\eeq
where $\cal{D}$ is the second order $q$-difference
operator \cite {ask}
having the Askey-Wilson polynomials $p_n$ as
eigenfunctions.
It is a linear transformation given by
\bea
  \mathcal{D}f[y] = (1+abcdq^{-1})f[y]
 &+&  \frac {(1-ay)(1-by)(1-cy)(1-dy)}{(1-y^2)(1-qy^2)}(f[qy]-f[y])
  \\  \nn
 &+&\frac {(a-y)(b-y)(c-y)(d-y)}{(1-y^2)(q-y^2)}(f[q^{-1}y]-f[y])
\label{54}
\eea
with $\mathcal{D}(1)=1+abcdq^{-1}$. The eigenvalue equation  for the
joint eigenfunctions $p_n$ reads
\beq
  \mathcal{D}p_n=\lambda_n^* p_n, \qquad \lambda^*_n = q^{-n}+abcdq^{n-1}
\label{55}
\eeq
and the operator $A^*$ is represented by an infinite-dimensional
matrix diag$(\lambda^*_0, \lambda^*_1, \lambda^*_2,...)$.
The operator $Ap_n=xp_n$ is represented by a tridiagonal matrix.
Let $\mathcal{A}$ denote the matrix whose matrix
elements enter the
three-term recurrence relation for the Askey-Wilson
polynomials
\beq
 xp_n=b_np_{n+1}+a_np_n+c_np_{n-1}, \qquad p_{-1}=0
\label{56}
\eeq
\beq
\mathcal{A} =
\begin{pmatrix} a_0 & c_1         \\
                b_0 & a_1 &  c_2   \\
                    & b_1 &  a_2 &\cdot   \\
                     &    &  \cdot & \cdot
\end{pmatrix}
\label{57}
\eeq
The explicit form of the matrix elements of $A$ reads
\beq
a_n=a+a^{-1}-b_n-c_n
\label{58}
\eeq
\beq
b_n=\frac {(1-abq^n)(1-acq^n)(1-adq^n)(1-abcdq^{n-1})}
{a(1-abcdq^{2n-1})(1-abcdq^{2n})}
\label{59}
\eeq
\beq
c_n=\frac {a(1-q^n)(1-bcq^{n-1})(1-bdq^{n-1})(1-cdq^{n-1})}
 {(1-abcdq^{2n-2})(1-abcdq^{2n-1})}
\label{60}
\eeq
The basis is orthogonal with the orthogonality condition
for the Askey-Wilson polynomials \cite {ask}
$P_n=a^{-n}(ab,ac,ad; q)_np_n$
\beq
\int_{-1}^1 \frac {w(x)}{2\pi \sqrt {1-x^2}}P_m(x; a,b,c,d\vert q)
 P_n(x; a,b,c,d\vert q)dx = h_n \delta_{mn}
\label{61}
\eeq
where $w(x)=\frac {h(x,1)h(x,-1)h(x,q^{1/2})h(x,-q^{1/2})}
 {h(x,a)h(x,b)h(x,c)h(x,d)}$ and $h(x,\mu)=\prod _{k=0}^{\infty}
[1-2\mu xq^k+\mu^2q^{2k}]$,
and
\beq
h_n=\frac {(abcdq^{n-1};q)_n(abcdq^{2n};q)_{\infty}}
{(q^{n+1}, abq^n,acq^n,adq^n,bcq^n,bdq^n,cdq^n;q)_\infty}
\label{62}
\eeq

As noted in the previous section a tridiagonal pair of operators $A, A^*$
is determined up to the affine transformation. One can
appropriately rescale the operators to obtain the algebraic relations
in the form
\bea
 [A, A^2A^*- \beta AA^*A +A^*A^2 -(q-q^{-1})^2 A^*]= 0  \\  \nn
 [A^*, A^{*2}A - \beta A^*AA^*+AA^{*2}-(q-q^{-1})^2A]= 0
\label{63}
\eea
which results in the corresponding rescaling of the
matrix elements. A shift of the operators has no other
effect but shifting the diagonal elements of the representing
matrices.

\section{Askey-Wilson algebra and reflection equation}

We consider models of statistical physics in which the spin
variable is associated with the site $i$ of a one-dimensional
lattice. An example of a model with quantum affine symmetry
is the spin $1/2$ XXZ model with Hamiltonian
defined on an infinite dimensional chain
\beq
   H=-\frac {1}{2}\sum_i (\sigma^x_i \sigma^x_{i+1}+
  \sigma^y_i\sigma^y_{i+1}+\Delta \sigma ^z_i\sigma^z_{i+1})
\label{64}
\eeq
where the Pauli matrices $\sigma^x_i,\sigma^y_i,\sigma^z_i$
act on the $i$-th component of the infinite tensor product
$...\otimes V_{i-1}\otimes V_i \otimes V_{i+1}\otimes...$,
with $V=\textbf{C}^2$. This model is known to be integrable \cite {jimmi}
within the representation theory of the affine quantized
algebra $U_q(\hat {sl}(2))$. Namely, given the $U_q(\hat {sl}(2))$
$R$-matrix operator $R(z_1/z_2)\in End_{\textbf{C}} V_{z_1}\otimes V_{z_2}$,
where $V_z$ is the two-dimensional $U_q(\hat {sl}(2))$ evaluation module,
satisfying the Yang-Baxter equation
\beq
R_{12}(z_1/z_2)R_{13}(z_1)R_{23}(z_2)=R_{23}(z_2)R_{13}(z_1)R_{12}(z_1/z_2)
\label{65}
\eeq
then, the Hamiltonian is written as $H=\sum H_{ii+1}$, where the
two site Hamiltonian density is obtained as
\beq
H_{ii+1}=\frac {d}{du}PR_{i i+1}\vert _{u=0}
\label{66}
\eeq
with $P$ the permutation operator and $z_1/z_2= e^u$.
The generators act on the quantum space by means of the infinite
coproduct and the invariance with respect to the affine $U_q(\hat {sl}(2))$
manifests in the property
\beq
 [H, \Delta^{\infty}(G_k)]=0
\label{67}
\eeq
for any of the generators $G_k$ of $U_q(\hat {sl}(2))$.
If we introduce for finite chain a boundary of  a particular form,
such as diagonal boundary terms, the symmetry is reduced to $U_q(sl(2))$
and the invariant Hamiltonian has the form \cite{sal}
\beq
H^{QGr}_{XXZ}=-1/2\sum_{i=1}^{L-1} (\sigma^x_i \sigma^x_{i+1}
+ \sigma^y_i \sigma^y_{i+1} + \Delta_q \sigma^z_i \sigma^z_{i+1}
 + h(\sigma^z_{i+1}-\sigma^z_i)+\Delta_q )
\label{68}
\eeq
where
\beq
\Delta_q=-\frac {1}{2}(q+q^{-1}), \quad h=\frac {1}{2}(q-q^{-1})
\label{69}
\eeq
In the presence of a boundary in addition to the
$R$-matrix there is one more matrix $K(z)$
which satisfies the boundary Yang-Baxter equation,
also known as a reflection equation.
\beq
R(z_1/z_2)(K(z_1)\otimes I)R(z_1z_2)(I\otimes K(z_2))
-(I\otimes K(z_2))R(z_1z_2)(K(z_1)\otimes I)R(z_1/z_2)=0
\label{69}
\eeq
Within the quantum inverse scattering method  the $K$-matrix is related to the
quantum current $L=L^+(L^-)^{-1}$ (10) where $L^{\pm} \in End V \otimes U_q(\hat {sl}(2))$.
In section (2) the two generators of the Askey-Wilson
algebra were constructed as linear covariant objects with the coproduct properties
of two-sided coideals of the quantum affine symmetry $U_q(\hat {sl}(2))$.
It is suggestive to construct the $K$-matrix in terms of the AW algebra
generators.

Let $R(z)$ be the symmetric trigonometric $R$-matrix with deformation parameter $q^{1/2}$
 \beq
 R(z) =
\begin{pmatrix} q^{1/2}z-q^{-1/2}z^{-1} &  0 & 0 & 0     \\
                                0  &   z-z^{-1}     &  q^{1/2}-q^{-1/2} &  0\\
                                0   & q^{1/2}-q^{-1/2} & z-z^{-1} & 0  \\
                                 0 & 0 & 0 &    q^{1/2}z-q^{-1/2}z^{-1}
\end{pmatrix}
\label{73}
\eeq
acting on the auxiliary tensor product space $V_{z_1} \otimes V_{z_2}$ which
carry the fundamental representations of the covariance algebra.  Then
one can construct an operator $L(z)$ \cite {kul1} in terms of the $U_q(sl(2))$
generators
\beq
 L(z) = \begin{pmatrix} zq^{J_3}-z^{-1}q^{-J_3} & (q^{1/2}-q^{-1/2})J_-         \\
                (q^{1/2}-q^{-1/2})J_+ &  zq^{-J_3}-z^{-1}q^{J_3}
\end{pmatrix}
\label{72}
\eeq
acting on the tensor product $V_0 \otimes V_Q$ of the auxiliary space $V_0$
and the quantum space $V_Q$ which in the general case carry  finite
dimensional inequivalent  $U_q(sl(2))$ representations.
The $L$-operator satisfies
\beq
  R(z_1/z_2)L_1(z_1)L_2(z_2)=L_2(z_2)L_1(z_1)R(z_1/z_2)
\label{73}
\eeq
where $L_1= L\otimes I$ and $L_2=I\otimes L$.
As it is known this relation together with the reflection equation (59)
constitute the basic algebraic
relations of the inverse scattering method to integrable models.

We are now going to construct a solution to eq.(59) in terms of the operators $A,A^*$.

{\it Proposition III}: Let $A, A^*$ generate the  AW algebra, the linear covariance algebra
for $U_q(sl(2))$. Then there exists a reflection matrix $K(z)=K^{op}(z)+K^c(z)$,
constructed in terms of the AW algebra generators,
where the part $K^{op}$ has the form
\beq
 K^{op}(z) =
\begin{pmatrix} q^{1/2}zA-q^{-1/2}z^{-1}\frac {\sqrt {\rho}}{\sqrt {\rho^*}}A^* &
                     -\frac {\sqrt {\rho}}{\sqrt {\rho^*}}(q^{1/2}-q^{-1/2})[A^*,A]_q        \\
               -\rho^{-1} \frac {\sqrt {\rho}}{\sqrt {\rho^*}}(q^{1/2}-q^{-1/2})[A,A^*]_q &
       -q^{-1/2}z^{-1}A+q^{1/2}z\frac {\sqrt {\rho}}{\sqrt {\rho^*}}A^*
\end{pmatrix}
\label{73}
\eeq
and the  part $K^c(z)$ is
\beq
K^c_{11}=\frac {q^{1/2}z\eta^*-q^{-1/2}z^{-1}\eta}{\rho(q^{1/2}+q^{-1/2})},
\quad\quad\quad K^c_{22}=\frac {q^{-1/2}z\eta-q^{1/2}z^{-1}\eta^*}{\rho(q^{1/2}+q^{-1/2})}
\label{65}
\eeq
$$K^c_{12}=-\rho \frac {q^{1/2}z^2+q^{-1/2}z^{-2}}{q^{1/2}+q^{-1/2}}
- \frac {\sqrt {\rho}}{\sqrt {\rho^*}}\omega, \quad\quad
K^c_{21}=-\frac {q^{1/2}z^2+q^{-1/2}z^{-2}}{q^{1/2}+q^{-1/2}}
- \rho^{-1} \frac {\sqrt {\rho}}{\sqrt {\rho^*}}\omega$$
The matrix $K(z)$ is a solution of the boundary Yang-Baxter equation (59) provided the operators
$A,A^*$ obey the tridiagonal algebraic relations of the AW algebra in the reduced general form (30)
with all structure constants $\rho, \rho^*, \omega, \eta, \eta^*$ nonzero.
We denote this solution $K(z, \rho)$.

The proof of this proposition is rather long but straightforward. It is directly verified
using the explicit form of the $R$ matrix (60) and the AW algebraic relations (30)
that the boundary matrix $K$ from (63, 64) solves the reflection equation (59).

We emphasize on the factor $\frac {\sqrt {\rho}}{\sqrt {\rho^*}}$ to  $A^*$ and $\omega$ in the $K$ matrix.
This factor is due to the fact that the solution of the boundary Yang-Baxter equation (59) in terms of
the AW algebra generators requires $\rho=\rho^*$. This is not a problem since given the AW algebra
in the general form (30) we can relate it to an algebra with $\rho=\rho^*$ rescaling
$A^* \rightarrow \frac {\sqrt {\rho}}{\sqrt {\rho^*}}A^*$. Alternatively we can rescale
$A \rightarrow \frac {\sqrt {\rho^*}}{\sqrt {\rho}}A$ to obtain an AW algebra with $\rho=\rho^*$.
This gives a second solution $K(z, \rho^*)$ of the reflection equation. Its
$K^{op}(z, \rho^*)$ part has the form
\beq
 K^{op}(z) =
\begin{pmatrix} q^{1/2}z\frac {\sqrt {\rho^*}}{\sqrt {\rho}}A-q^{-1/2}z^{-1}A^* &
                     -\frac {\sqrt {\rho^*}}{\sqrt {\rho}}(q^{1/2}-q^{-1/2})[A^*,A]_q        \\
               -\rho^{*-1} \frac {\sqrt {\rho^*}}{\sqrt {\rho}}(q^{1/2}-q^{-1/2})[A,A^*]_q &
       -q^{-1/2}z^{-1}\frac {\sqrt {\rho^*}}{\sqrt {\rho}}A+q^{1/2}zA^*
\end{pmatrix}
\label{73}
\eeq
The matrix elements of the $K^c(z,\rho^*)$ part are obtained from (64) by the interchange
$\rho \leftrightarrow \rho^*$. The solution $K(z, \rho^*)$ can be implemented
to construct a solution $K^*(z)$ of the dual reflection equation \cite {skl2, veg}
Namely, the matrix  $K^*(z) = K^t(z^{-1},\rho^*)$ solves the dual reflection equation,
(which is obtained from eq.(59) by changing
$z_{1,2}\rightarrow q^{-1/2}z_{1,2}^{-1}$ and $K\rightarrow K^t$).

Setting $\rho=\rho^*$ and $\eta =\eta^* = 0$ in (63) and (64) we obtain the $K$ matrix
considered in \cite {bas2} for  such a  very particular case of an AW algebra and for the
spin 1/2 quantum space representation. We note that an AW algebra in the general form with a
sequence of scalars
 $-(q+q^{-1}), \gamma, \gamma^*, \omega, \eta, \eta^*$
cannot be reduced to such a  particular algebra with
structure constants $-(q+q^{-1}), \rho, \rho, 0,0, \omega, 0,0$.
There exists an unique affine transformation \cite {vid2} to only set
$\gamma=\gamma^*=0$ (and simultaneously either $\rho=\eta^*=0$ or $\eta=\rho^*=0$).

\section{A model of nonequilibrium physics with boundary Askey-Wilson algebra}

Reaction-diffusion processes provide a playground to increase the utility of quantum
groups \cite {kul3}. As a physical example we consider the asymmetric simple
exclusion process (ASEP), a model of nonequilibrium physics with rich behaviour
and wide range of applicability \cite {schu, schre, mac, spo}.

The asymmetric exclusion process is an exactly solvable model of a
lattice diffusion system of particles interacting with a hard core
exclusion, i.e. the lattice site can be either empty or occupied
by a particle. As a stochastic process it is described in terms of
a probability distribution $P(s_i, t)$ of a stochastic variable
$s_i = 0, 1$ at a site $i = 1,2,....L$ of a linear chain. A state
on the lattice at a time $t$ is determined by the occupation
numbers $s_i$ and a transition to another configuration $s_i'$
during an infinitesimal time step $dt$ is given by the probability
$\Gamma(s, s')dt$. Due to probability conservation $\Gamma (s,s) =
-\sum_{s'\neq s} \Gamma (s',s)$. The rates $\Gamma \equiv \Gamma
^{ik}_{jl}$, $i,j,k,l =0, 1$ are assumed to be independent from
the position in the bulk. For diffusion processes the transition
rate matrix becomes simply $\Gamma ^{ik}_{ki}=g_{ik}$.  At the
boundaries, i.e. sites $1$ and $L$ additional processes can take
place with rates $L_i^j$ and $R_i^j$ ($i,j =0, 1$). In the set of
occupation numbers $(s_1,s_2,...,s_L)$ specifying a configuration
of the system $s_i=0$ if a site $i$ is empty, $s_i=1$ if there is
a particle at a site $i$. Particles hop to the left with
probability $g_{01}dt$ and to the right with probability
$g_{10}dt$. The event of exchange happens if out of two adjacent
sites one is a vacancy and the other is occupied by a particle.
The symmetric simple exclusion process is known as the lattice gas
model of particles hopping between nearest-neighbour sites with a
constant rate $g_{01}=g_{10}=g$. The partially asymmetric simple
exclusion process with hopping in a preferred direction is the
driven diffusive lattice gas of particles moving under the action
of an external field. The process is totally asymmetric if all
jumps occur in one direction only, and partially asymmetric if
there is a different non-zero probability of both left and right
hopping. The number of particles in the bulk is conserved and this
is the case of periodic boundary conditions. In the case of open
systems, the lattice gas is coupled to external reservoirs of
particles of fixed density. Phase transitions inducing boundary
processes \cite {kru} are the most interesting examples  (see
\cite {der2} for a review) when a particle is added with
probability $\alpha dt$ and/or removed with probability $\gamma
dt$ at the left end of the chain, and it is removed with
 probability
$\beta dt$ and/or added with probability $\delta dt$
 at the right
 end of the chain. Without
loss of generality we can choose the right
probability
rate $g_{10}=1$ and the left probability rate
$g_{01}=q$.

The time evolution of the model is governed
by the master equation for the
probability distribution of the stochastic system
\beq
 \frac {dP(s,t)}{dt} = \sum _{s'} \Gamma (s,s')P(s',t)
\eeq \label{1}It can be mapped to a Schroedinger equation in
imaginary time for a quantum Hamiltonian with nearest-neighbour
interaction in the bulk and single-site boundary terms \beq
  \frac {dP(t)}{dt}=-HP(t)
\eeq \label{2}where$H= \sum_j H_{j,j+1}+ H^{(L)} + H^{(R)}$. The
ground state of this, in general non-Hermitian, Hamiltonian
corresponds to the stationary probability distribution of the
stochastic dynamics. The mapping provides a connection to the
integrable $SU_q(2)$-symmetric $XXZ$ quantum spin chain with
anisotropy $\Delta = \frac {(q+q^{-1})}{2}$, $q=\frac
{g_{01}}{g_{10}} \neq 1$ and most general boundary terms.

We consider the model within the matrix-product-state ansatz of
stochastic dynamics \cite {der2, der1}, which was inspired by the
inverse scattering method to integrable models. The idea is that
one associates with an
occupation number $s_i$ at a position $i$ a matrix
$D_{s_i}=D_1$ if a site $i=1,2,...,L$ is occupied and
$D_{s_i}=D_0$ if a site $i$ is empty and
the
stationary probability distribution is expressed
as a product of (or a trace over) matrices
that form a representation of a quadratic algebra
\beq
D_1D_0-qD_0D_1=x_0D_1-D_0x_1, \qquad x_0+x_1=0
\label{4}
\eeq
where $0<q<1$ and $x_0, x_1$ are representation dependent
parameters. The totally asymmetric process corresponds to $q=0$
and the symmetric process - to $q=1$.
The quadratic algebra with no $x$-terms on the RHS
(i.e. $D_1D_0-qD_0D_1=0$)
corresponds to a bulk process with reflecting boundaries.

For an open system with boundary processes the normalized
steady weight of a given configuration is expressed as a
matrix element in an auxiliary vector space
\beq
 P(s_1,....s_L)=\frac {\langle w\vert D_{s_1}D_{s_2}...D_{s_L}
   \vert v\rangle}{Z_L},
\eeq
\label{6}
with respect to the vectors $\vert v>$ and $<w\vert $,
defined by the boundary conditions
\bea
  (\beta D_1-\delta D_0)\vert v\rangle &=&x_0\vert v\rangle  \\  \nn
  \langle w\vert (\alpha D_0 - \gamma D_1)&=&\langle w\vert(-x_1)
\eea
\label{7}
The normalization factor to the stationary probability
distribution is
\beq
Z_L=\langle w \vert (D_0+D_1)^L \vert v\rangle
\label{8}
\eeq
In the following we set $x_1=-x_0$.

The advantage of the matrix-product ansatz is that once the
representation of the diffusion algebra and the boundary
vectors $\vert v>$ and $<w\vert $ are known, one can
evaluate all the relevant physical quantities such as
the mean density  at a site $i$, $ \langle s_i\rangle =\frac {\langle
   w\vert (D_0+D_1)^{i-1}D_1(D_0+D_1)^{L-i}
   \vert v\rangle}{Z_L} $,
the two-point correlation function
$\langle s_is_j\rangle =\frac {\langle w\vert
  (D_0+D_1)^{i-1}D_1(D_0+D_1)^{j-i-1}D_1
  (D_0+D_1)^{L-j}\vert v\rangle}{Z_L}$
and higher correlation functions.
The current $J$ through a bond between site $i$
and site $i+1$, $J= \frac {\langle
   w\vert (D_0+D_1)^{i-1}(D_1D_0-qD_0D_1)(D_0+D_1)^{L-i-1}
   \vert v\rangle}{Z_L} $
has a very simple form  $J=x_0 \frac {Z_{L-1}}{Z_L} $. The matrix
 $D_0+D_1$ enters all the expressions and plays the role
 of a transfer matrix operator.

The algebraic matrix state approach (MPA) is the equivalent
formulation of recursion relations derived for the asymmetric
exclusion process (ASEP)
in earlier works \cite {do, san} which could not be  readily
generalized to other models. In most applications one uses
infinite dimensional representations of the quadratic algebra.
Finite dimensional representations \cite {ess, ma}
impose a constraint on the model parameters. They may be useful in relation
to Bethe Ansatz on a ring \cite {al}. The MPA was
generalized to many-species models
\cite {der2, ar, r} and to dynamical MPA \cite {sti}.

  The matrices $D_0, D_1$  of the MPA generate an AW algebra with
 $\rho=\rho^*=0$ \cite {an}.
We call this algebra the bulk Askey-Wilson algebra. For the particular
case of only incoming (outgoing) particle at left (right) end of the chain
the boundary operators satisfy an isomorphic $\rho=\rho^*=0$ AW algebra
which can be solved
by   shifted  $q$-deformed oscillators \cite{co, chai, ku}
as they
were applied for the ASEP with such particular boundary conditions
 \cite{ev, sa}.
 In the general case of incoming and
outgoing particles at both boundaries
there are four operators $\beta D_1, -\delta D_0,
-\gamma D_1, \alpha D_0$
and one needs an addition rule to form two linear
independent boundary
operators acting on the dual boundary vectors.
From the quadratic algebra (68)
 two independent relations for the boundary operators  follow
\beq
\beta D_1\alpha D_0-q\alpha D_0\beta D_1=
 x_0 (\alpha \beta D_1+\beta \alpha D_0)
\label{19}
\eeq
and
\beq
\gamma D_1 \delta D_0-q\delta D_0\gamma D_1=
x_0(\delta \gamma D_1+\gamma \delta D_0)
\label{20}
\eeq
To form two linearly independent operators
$B^R=\beta D_1-\delta D_0, B^L=-\gamma D_1 +\alpha D_0$
for a  solution to the  boundary problem and
in order to emphasize
the equivalence of the ASEP to the integrable spin $1/2$ $XXZ$
one can use the $U_q(su(2))$ algebra.
It is generated by three
elements with the defining commutation relations
\beq
[N, A_{\pm}]= \pm A_{\pm} \qquad [A_+, A_-]=\frac {q^N-q^{-N}}{q^{1/2}-q^{-1/2}}
\label{22}
\eeq
and a central element
\beq
   Q=A_+A_- - \frac {q^{N-1/2} -q^{-N+ 1/2}}{(q^{1/2}-q^{-1/2})^2}
\label{23}
\eeq
The relations (72), (73)
can be solved by choosing a representation of the boundary operators
in the form
\bea
\beta D_1 -\delta D_0 &=&  \\  \nn
 \frac {x_0\beta}{\sqrt {1-q}}q^{N/2}A_+
 -\frac {x_0\delta}{\sqrt {1-q}}A_-q^{N/2}
  -x_0\frac{-\beta q^{1/2}+\delta}{1-q}q^N
  +x_0\frac {\beta -\delta}{1-q}    \\ \nn
\alpha D_0 - \gamma D_1 &=&  \\  \nn
\frac {x_0\alpha}{\sqrt {1-q}}q^{-N/2}A_+ -
\frac {x_0\gamma }{\sqrt {1-q}}A_-q^{-N/2}
+x_0\frac {\alpha q^{-1/2} -\gamma }{1-q}q^{-N} +
  x_0\frac {\alpha  -\gamma}{1-q}
\label{26}
\eea
Separating the shift parts from the boundary operators and denoting
the corresponding rest operator parts by $A$ and $A^*$ we write the
left and right boundary operators in the form
\bea
 \beta D_1 -\delta D_0 &=& A  +x_0\frac {\beta -\delta}{1-q}  \\ \nn
 \alpha D_0 -\gamma D_1 &=& A^* + x_0\frac {\alpha -\gamma}{1-q}
\label{27}
\eea
Then the operators $A$ and $A^*$ defined by
\bea
 A&=&
 \beta D_1 -\delta D_0  -x_0\frac {\beta -\delta}{1-q} \\ \nn
 A^*&=&
 \alpha D_0 - \gamma D_1 -x_0\frac {\alpha - \gamma}{1-q}
\label{28}
\eea
and their $q$-commutator
\beq
 [A,A^*]_q=q^{1/2}AA^*-q^{-1/2}A^*A
\label{29}
\eeq
satisfy the boundary Askey-Wilson algebra of the open ASEP
\bea
[[A,A^*]_q,A]_q &=& -\rho A^* -\omega A - \eta  \\ \nn
[A^*,[A,A^*]_q]_q &=& -\rho^*A -\omega A^* -\eta^*
\label{30}
\eea
with structure constants given by
\beq
  \rho=x_0^2 \beta \delta q^{-1}(q^{1/2}+q^{-1/2})^2, \qquad
  \rho^*=x_0^2 \alpha \gamma q^{-1}(q^{1/2}+q^{-1/2})^2
\label{31}
\eeq
\beq
 -\omega=x_0^2(\beta -\delta)(\gamma- \alpha)
    -x_0^2( \beta \gamma +\alpha \delta) (q^{1/2}-q^{-1/2})Q
\label{32}
\eeq
\bea
\eta&=&q^{1/2}(q^{1/2}+q^{-1/2})x_0^3\left (\beta \delta ( \gamma
 -\alpha)Q +\frac {(\beta -\delta)(\beta \gamma
 +\alpha \delta)}{q^{1/2} - q^{-1/2}} \right ) \\  \nn
\eta^*&=&q^{1/2}(q^{1/2}+q^{-1/2})x_0^3\left (\alpha \gamma (\beta
 -\delta)Q + \frac {( \alpha - \gamma)(\alpha \delta
 +  \beta \gamma)}{q^{1/2}-q^{-1/2}} \right )
\label{33}
\eea
One can further use  the affine transformation properties of
the AW algebra generators to obtain a representation of the
boundary ASEP operators from the basic representation of the AW algebra.
We summarize the results for the representation of the ASEP
boundary operators (for details see \cite {an} ):

There is a representation $\pi$ in a space with basis the AW polynomials
$p_n=p_n(x;a,b,c,d)$ (41)
\beq
   (p_0,p_1,p_2,...)^t
\label{68}
\eeq
with respect to which the right boundary operator
$D_1- \frac {\delta}{\beta} D_0 \equiv D_1+bdD_0$ is diagonal.
The representing  matrix is diag$(\lambda_0,\lambda_1,\lambda_2,...)$
with the eigenvalues $\lambda_n$ given by
\beq
  \lambda_n= \frac {q^{1/2}}{1-q}\left (bq^{-n} + dq^{n}
  +1+bd \right )
\label{69}
\eeq
The left boundary operator
$D_0 - \frac {\gamma}{\alpha}D_1 \equiv D_0 + acD_1$ is
tridiagonal and its representing matrix has the form
\beq
  \pi (D_0+acD_1) = \frac {q^{1/2}}{1-q} \left (b\mathcal{A}^t +1+ac \right )
\label{70}
\eeq
where the matrix $\mathcal{A}$ is given by (46).
The dual representation $\pi^*$
has a basis
\beq
  (p_0,p_1,p_2,...)
\label{71}
\eeq
with respect to which the left boundary operator
$\pi^* (D_0+acD_1)$ is diagonal
$diag(\lambda_0^*, \lambda_1^*,...)$
with diagonal elements
\beq
 \lambda^*_n=\frac {q^{1/2}}{1-q}\left (aq^{-n} + cq^{n}
  +1+ac \right )
\label{72}
\eeq
The right boundary operator is represented by a tridiagonal
matrix
\beq
\pi^* (D_1+bdD_0)= \frac {q^{1/2}}{1-q} \left (a\mathcal{A} +
1+bd \right)
\label{73}
\eeq

The ASEP boundary value problem is satisfied with the left and right
boundary vectors chosen in the form
\beq
  \langle w \vert = h_0^{-1/2}(p_0, 0,0,....)\qquad
 \vert v \rangle = h_0^{-1/2}(p_0, 0,0,...)^t
\label{69}
\eeq
where $h_0$ is a normalization from the orthogonality condition (50).
With this choice the solutions to the boundary eigenvalue equations
uniquely relate (in this representation) the four parameters
of the Askey-Wilson polynomials with the boundary probability rates
\beq
a=\kappa^*_+,\quad b=\kappa_+, \quad c=\kappa^*_-, \quad
 d=\kappa_-
\label{73}
\eeq
where
\bea
 \kappa_{\pm}&=& \frac {-(\beta -\delta -(1-q)) \pm
  \sqrt {(\beta -\delta -(1-q))^2 +4\beta \delta}}{2\beta} \\  \nn
 \kappa^*_{\pm}&=&\frac {-(\alpha - \gamma -(1-q)) \pm
   \sqrt {(\alpha - \gamma -(1-q))^2 +4\alpha \gamma}}{2\alpha}
\label{72}
\eea
The expressions on the RHS of eq.(92) are the functions of the parameters
which define the  phase diagram of the ASEP. They have been used in
previously known MPA applications where have always been
taken for granted. It is quite remarkable that here they follow from the
properties of the Askey-Wilson algebra representations.

It can be further shown that the transfer matrix $D_0+D_1$
and each of the boundary operators generate isomorphic
AW algebras \cite {an}. In the tridiagonal representation the
transfer matrix $D_0+D_1$ satisfies the
three-term recurrence relation of the AW polynomials which was
explored in \cite {wa} for the solution of the ASEP in
the stationary state. The exact
calculation of all the physical quantities, such as the current,
correlation functions etc,   in terms of
the Askey-Wilson polynomials was achieved without any reference to
the AW algebra. The ultimate relation of the  exact solution in
the stationary state  to the AW polynomials was possible due to
the  AW boundary hidden
symmetry of the ASEP with most general boundary conditions.
The relation of the AW algebra to the  $K$-matrix,
determined by (63) and (64), and satisfying the reflection equation,
puts  a  solution  beyond the stationary state into perspective.

\section{Interpretation of the ASEP boundary operators}

As known the open ASEP is related to the integrable spin
$1/2$ XXZ quantum spin chain
through the similarity transformation
$\Gamma =- q U^{-1}_{\mu}H_{XXZ}U_{\mu}$ (see \cite {ess} for details).
$H_{XXZ}$ is the Hamiltonian of the $U_q(su(2))$ invariant
quantum spin chain (57) with anisotropy $\Delta_q$ and with added
non diagonal boundary terms $B_1$ and $B_L$.
\beq
H_{XXZ}=H_{XXZ}^{QGr} + B_1 +B_L
\label{76}
\eeq
The transition rates of the ASEP are related to the
boundary terms in the following way
($\mu$ is a free parameter, irrelevant for the spectrum)
\bea
B_1&=&\frac {1}{2q}\left( \alpha + \gamma +
(\alpha - \gamma) \sigma^z_1 -2\alpha \mu \sigma_1^-
-2\gamma \mu^{-1} \sigma_1^+ \right)          \\  \nn
B_L&=&\frac {\left( \beta +\delta
-(\beta-\delta) \sigma^z_L -2\delta \mu q^{L-1}\sigma^-_L
-2\beta \mu^{-1} q^{-L+1}\sigma^+_L \right)}{2q}
\label{75}
\eea
It has been shown by
Sandow and Schuetz \cite{schu2} that the bulk driven diffusive system with
reflecting boundaries can be mapped to the spin $1/2$ $U_q(su(2))$-invariant quantum
spin chain. The $U_q(su(2))$ generators satisfying eqs.(74) and (75)  act on the
tensor product representation space $(V^2)^{\otimes L}$  as
\bea
q^{\pm N}=q^{\pm \frac {\sigma_3}{2}}\otimes q^{\pm \frac {\sigma_3}{2}}
  \otimes ... \otimes q^{\pm \frac {\sigma_3}{2}}  \\  \nn
A_{\pm}=\sum_i q^{\frac {\sigma_3}{4}}\otimes ... \otimes q^{\frac {\sigma_3}{4}}
\otimes \sigma_i^{\pm} \otimes
q^{-\frac {\sigma_3}{4}}\otimes ... \otimes q^{\frac {-\sigma_3}{4}}
\label{77}
\eea
where $\sigma_3, \sigma^{\pm}$ are the Pauli matrices and the index $i$ means that the matrix
is associated with the $i$th site of the chain ($i$th position in the tensor product).
The representation is completely reducible, the product of $L$ spin $1/2$ representations
decomposes into a direct sum of spin $j$ irreducible representations
with maximal highest weight $j=L/2$ decreasing by $1$ to $j=0$ or $j=1/2$ for even $L$
or odd $L$. Within the matrix product approach the bulk process with
reflecting boundary conditions is described by a quadratic algebra
\beq
 D_1D_0-qD_0D_1=0
\label{78} \eeq which defines a two-dimensional noncommutative
plane with the $SU_q(2)$ action as its symmetry. The operators
associated with the bulk ASEP form the two- dimensional comodule
of $SU_q(2)$. As a consequence of eq.(96), for generic $q$, a spin
$j$ representation of $U_q(su(2))$ can be realized in the space of
the $q$-symmetrized product of $L=2j$ two dimensional
representations $D_\mu$, $\mu=0,1$, with basis $D^{L-k}_0D^k_1$,
$k=0,1,...,L$. The stationary probability distribution, i.e. the
ground state of the $U_q(su(2))$ invariant Hamiltonian
$H_{XXZ}^{QGr}$, corresponds to the $q$-symmetrizer of the Young
diagram with one row and $L$ boxes \cite{alc}. The presence of the
boundary processes (i.e. the nondiagonal boundary terms in the
Hamiltonian) reduces the $U_q(su(2))$ bulk invariance and amounts
to the appearance of linear terms in the quadratic algebra. The
boundary conditions define the boundary operators which carry a
residual symmetry of the process. It is expressed in the fact that
the boundary operators are constructed in terms of the
$U_q(su(2))$ generators, as seen from the explicit formulae (76).
With $A_{\pm}, N$ being the generators of a finite dimensional
$U_q(su(2))$ representation, it can be verified from eq.(76) that
$\alpha D_0-\gamma D_1$ commutes with $H(q)^{QGr}$ and $\beta
D_1-\delta D_0$ commutes with $H(-q^{-1})^{QGr}$, where according
to \cite {sal} \beq H^{QGr}(-q^{-1})=-UH^{QGr}(q)U^{-1} \label{79}
\eeq and \beq
 U=\exp \left (i\frac {\pi}{2} \sum_{m=1}^L m\sigma_m^3 \right )
\label{80}
\eeq
Thus the boundary operators constructed as the linear covariant objects of the
bulk $U_q(su(2))$ symmetry acquire a very important physical meaning - they can
be interpreted as the two nonlocal conserved charges of the open ASEP. Such non
local boundary symmetry charges were originally obtained for the sine Gordon
model \cite{nep} and generalized to affine Toda field theories \cite{del}, and
derived from spin chain point of view as commuting with the transfer matrix for
a special choice of the boundary conditions \cite{doi}.
In particular, the left boundary
operator $\alpha D_0-\gamma D_1$ in the finite dimensional representation (76) is
analogous to the one boundary Temperley-Lieb algebra centralizer in the "nondiagonal"
spin $1/2$ representation \cite{nic}.

\section{Discussion and conclusion}

In this paper we have considered the homomorphism of the Askey-Wilson
algebra with two generators into the  quantum affine $U_q(\hat {sl}(2))$ algebra.
This homomorphism defines the Askey-Wilson algebra as a coideal subalgebra of
$U_q(\hat {sl}(2))$.  We have constructed an AW operator valued $K$ matrix which is
a solution of the boundary Yang-Baxter equation (reflection equation). We consider
the relation of an AW algebra to a solution of the reflection equation to be
important for the exact solvability of a physical system with  the quantum
$U_q(\hat {sl}(2))$ invariance in the bulk and hidden boundary Askey-Wilson
algebra symmetry.

As an example of a physical system with boundary AW algebra we
consider a model of nonequilibrium physics, the open asymmetric
exclusion process with general boundary conditions. This model is
equivalent to the integrable spin $1/2$ $XXZ$ chain with
nondiagonal boundary terms whose bulk invariance (infinite spin
chain) is $U_q(\hat {su}(2))$. The presence of  boundaries breaks
the bulk quantum affine symmetry of the equivalent quantum spin
chain, however a remnant of the bulk symmetry survives and it is
expressed in the possibility to construct the ASEP boundary
operators in terms of the $U_q(\hat {su}(2))$ generators. Thus the
boundary operators of the open asymmetric exclusion process
generate an Askey-Wilson algebra, which is the hidden boundary
symmetry of the process. The exact solution of the ASEP with most
general boundary conditions (four boundary probability rates) in
the stationary state was obtained \cite{wa} in terms of the AW
polynomials without reference to the AW algebra. It was emphasized
that the solution was  ultimately related with the AW polynomials.
Such an ultimate relationship is natural from the point of view of
the boundary AW algebra. The existence of the reflection matrix
$K(z, \rho)$ (and its dual $K^t(z^{-1},\rho^*)$) constructed in
terms of the AW algebra generators and  satisfying the boundary
Yang-Baxter equation is, in our opinion, {\it the deep algebraic
property of the open asymmetric exclusion process that may allow
for extending its exact solvability beyond the stationary state}.

It is important to emphasize the representation dependence of the Askey-Wilson algebra
(as well as of the MPA bulk quadratic algebra (68)). Constructed as a coideal subalgebra,
it has the property that the
structure constants $\rho, \rho^*, \omega, \eta, \eta^*$  carry the information
of the corresponding quantum algebra $U_q(\hat su(2))$.
The boundary Askey-Wilson algebra whose structure constants depend on the finite
dimensional $U_q(\hat su(2))$ representations is the ASEP hidden symmetry and this may
have  an important consequence in relation to Bethe Ansatz integrability.
The Bethe solution of the open ASEP \cite{deg1} was achieved through the mapping to the
$U_q(su(2))$ integrable $XXZ$ quantum spin chain with most general non diagonal boundary terms,
provided a particular constraint on the model parameters was satisfied. Quite surprisingly
the constraint coincides with the condition for a finite-dimensional representation
of the Askey-Wilson boundary algebra. The suitably chosen representation dependent boundary algebra
may turn to be the key in relation to Bethe Ansatz integrability.
For the ASEP the  reduction of the bulk invariance
gives rise to the boundary
symmetry which remains as the linear covariance algebra of the bulk $U_q(\hat su(2))$ symmetry
and  one can further employ Bethe
ansatz to obtain exact results for the approach to stationarity at large times and to
completely determine the spectrum of the transfer matrix. As commented in \cite {deg2} the
way one can satisfy the condition for the Bethe Ansatz solution of the ASEP implies additional
symmetries. In our opinion, the linear covariance Askey-Wilson algebra of the bulk $U_q(\hat su(2))$,
whose generators are interpreted as the two nonlocal conserved charges of the ASEP, is the hidden
symmetry behind Bethe Ansatz solvability.

It is worth mentioning that the relation of the ASEP (or the equivalent quantum spin
chain) boundary algebra to Bethe Ansatz integrability is promising form the point of view
of  Bethe Ansatz perspective in string theory. One is interested in closed strings with periodic
boundary conditions. However,  it is simpler to find the scattering matrix  on the infinite line
using asymptotic states and bootstrap. Then the spectrum is determined by asymptotic
Bethe equations \cite {art, sta} and they are approximate for a system of a finite size. The study
of the Askey-Wilson  algebra of a system on a ring with periodic boundary conditions  which is
interesting on its own, might be also useful for application to strings of finite length.

We have obtained an Askey-Wilson algebra  as a coideal subalgebra of the
quantum affine $U_q(\hat {sl}(2)$ and implemented it to find a solution of the reflection equation.
We have related this consideration to a model of nonequilibrium physics where the
 boundary operators generate a tridiagonal Askey-Wilson algebra, which is the linear
covariance algebra of the bulk $U_q(\hat su(2))$ symmetry. It is the hidden symmetry that
allows for the exact solvability in the stationary state and provides the framework
for employing Bethe ansatz to determine the dynamical properties of the open process.

\section{Acknowledgments}

The authors are grateful to V.Rittenverg and R.B.Zhang for
interest in the work and stimulating discussions. The comments of
P.Baseilhac and P.Terwilliger are appreciated. B.A. and P.P.K.
would like to thank the Department of Physical Sciences,
University of Helsinki, and the Helsinki Institute of Physics for
the hospitality during their visits in September-October 2007 when
this work was completed. The support of the Academy of Finland and
the Bulgarian Academy of Sciences under  a joint research  project
is acknowledged.

\end{document}